\documentclass{amsart}

\usepackage[dvips]{graphicx}
\usepackage[b5paper,twoside,top=20mm,right=18mm,left=22mm,bottom=15mm,bindingoffset=0mm,nomarginpar]{geometry}
\usepackage{amssymb,latexsym,tensor}
\usepackage{eqnarray,amsmath}
\usepackage{mathtools}

\newtheorem{theorem}{Theorem}[section]
\newtheorem{definition}{Definition}[section]
\newtheorem{conjecture}{Conjecture}[section]

\newtheorem{remark}{Remark}[section]

\numberwithin{equation}{section}

\newcommand{\rmd}{\textrm d}
\author[V. Pasic, \ E. Barakovic, \ N. Okicic]{Vedad Pasic, \ Elvis Barakovic, \ Nermin Okicic}
\address{
Department of Mathematics \newline \indent
University of Tuzla \newline \indent
Bosnia and Herzegovina}
\email{vedad.pasic@untz.ba, elvis.barakovic@untz.ba, nermin.okicic@untz.ba}

\title[\uppercase{A New Representation of the QMAG Field Equations}]{\uppercase{A New Representation of the Field Equations of Quadratic Metric--Affine Gravity}}


\begin{document}
\begin{abstract}
We deal with quadratic metric--affine gravity (QMAG), which is an alternative theory
of gravity and present a new explicit representation of the field equations of this
theory. In our previous work we found new explicit vacuum solutions of QMAG, namely generalised
pp-waves of parallel Ricci curvature with purely tensor torsion.
Here we do not make any assumptions on the properties of torsion and write down our field equations
accordingly. We present a review of research done thus far by several authors
in finding new solutions of QMAG and different approaches in generalising pp-waves.
We present two conjectures on the new types of solutions of QMAG
which the ansatz presented in this paper will hopefully enable us to prove.
\end{abstract}
\maketitle

\section{Introduction}
In 1905, \emph{Albert Einstein} published his work on the theory of \emph{special relativity}.
Classical mechanics and classical electromagnetism provide models that are good representations
of two sets of actual experiences. As Einstein noted in \cite{einsteinSR},
it is not possible to combine these into a single self--consistent model.
The construction of the simplest possible self--consistent model by Einstein
is the achievement of Einstein's theory of special relativity. Special relativity
gave a very satisfactory representation of the electromagnetic
interaction between charged particles, but the theory itself does not
deal with gravitational interaction.

General relativity is a theory of gravitation that was developed by
Einstein between 1907 and 1915. Hermann Minkowski put Einstein's special relativity
model into geometrical terms, and it is widely believed that Einstein
constructed his theory of general relativity by experimenting with
the generalisation of the geometric model.

Two problems with general relativity became apparent quite quickly.
Einstein considered that what are recognised locally as inertial
properties of local matter must be determined by the properties
of the rest of the universe. To what extent general relativity
manages to do this is still unclear to this day, although
Einstein's efforts to discover this extent founded
the modern study of \emph{cosmology}. The second problem of general relativity was that, although electromagnetism pointed the way
to general relativity, it is not included in the theory itself. As is evident from his remarks in  \cite{Einstein Meaning of Relativity},
Einstein expected much more from general relativity
than `just' the amalgamation of gravitation and electromagnetism at the macroscopic level.
He thought the theory should explain the existence of elementary particles and should
provide a treatment for nuclear forces. He spent most of the second part
of his life in pursuit of this aim, but with no real success.

There are a number of different \emph{alternative} theories of gravity that try to
further the completion of Einstein's theory of gravity.
One such theory, propagated by Einstein himself for some time,
is \emph{metric--affine gravity}, which is the theory employed by
this paper. Metric--affine gravity is a natural generalisation of Einstein's general relativity,
which is based on a spacetime with a Riemannian metric $g$ of Lorentzian signature. In metric--affine gravity, we consider spacetime to be a connected real $4$-manifold $M$ equipped with a Lorentzian
metric $g$ and an affine connection $\Gamma$. The $10$ independent components of the symmetric
metric tensor $g_{\mu\nu}$ and the $64$ connection coefficients $\tensor{\Gamma}{^\lambda_{\mu\nu}}$ are the unknowns of our theory, see \cite{hehlreview} for more details.

\begin{definition}\label{riemann}We call a spacetime $\{M,g,\Gamma\}$ Riemannian if the connection is Levi--Civita (i.e.
$\tensor{\Gamma}{^\lambda_{\mu\nu}}=\left\{{\lambda\atop {\mu\nu}}\right\}$), and non--Riemannian otherwise.
\end{definition}

The spacetime of metric--affine gravity reduces to that of general relativity provided
that the torsion (\ref{torsion}) of the connection $\Gamma$ vanishes and that the connection is metric
compatible (i.e. the covariant derivative of the metric $g$ vanishes, $\nabla g\equiv 0$). In this case
the connection is uniquely determined by the metric (Levi--Civita connection) and the
same is true for the curvature. Consequently, the metric $g$ is the only unknown quantity
of Einstein's equation. In contrast, the metric--affine approach does not involve any a
priori assumptions about the connection $\Gamma$ and thus the metric $g$ and the connection $\Gamma$
are viewed as two totally independent unknown quantities.

In \emph{quadratic} metric--affine gravity (QMAG), we define our action as
\begin{equation}\label{action}
S:=\int q(R)
\end{equation}
where $q$ is an $O(1, 3)$--invariant quadratic form on curvature $R$. Independent variation
of the metric $g$ and the connection $\Gamma$ produces Euler--Lagrange equations which we will
write symbolically as
\begin{align}
\label{variation-S-metric}
\partial S/\partial g&:=0\\
\label{variation-S-connecion}
\partial S/\partial \Gamma&:=0.
\end{align}

Our objective is the study of the combined system of field equations (\ref{variation-S-metric}), (\ref{variation-S-connecion}). This is a system of $10+64$ real nonlinear partial differential equations with $10+64$ real unknowns. The quadratic form $q(R)$ has $16$ $R^2$ terms with $16$ real coupling constants, and it can be represented as
\begin{multline}
\label{qfmain} q(\!R)=b_1\mathcal{R}^2+b^*_1\mathcal{R}_*^2
\\
+\sum_{l,m=1}^3 b_{6lm}(\mathcal{A}^{(l)},\mathcal{A}^{(m)})
+\sum_{l,m=1}^2 b_{9lm}(\mathcal{S}^{(l)},\mathcal{S}^{(m)})
+\sum_{l,m=1}^2 b^*_{9lm}(\mathcal{S}_*^{(l)},\mathcal{S}_*^{(m)})
\\
+b_{10}(R^{(10)},R^{(10)})_{\mathrm{YM}}
+b_{30}(R^{(30)},R^{(30)})_{\mathrm{YM}}
\end{multline}
with some real constants $b_1$, $b^*_1$, $b_{6lm}=b_{6ml}$,
$b_{9lm}=b_{9ml}$, $b^*_{9lm}=b^*_{9ml}$, $b_{10}$, $b_{30}$. Here
$\mathcal{R}$, $\mathcal{R}_*$, $\mathcal{A}^{(l)}$,
$\mathcal{S}^{(l)}$, $\mathcal{S}_*^{(l)}$, $R^{(10)}$, $R^{(30)}$
are tensors representing the irreducible pieces of curvature
and the inner products
$(\cdot,\cdot)$ and $(\cdot,\cdot)_{YM}$ are defined by
\begin{equation*}
(K,L):=K_{\mu\nu}\,L^{\mu\nu}\,,
\quad \quad
 (R,Q)_{\mathrm{YM}}:=R^\kappa{}_{\lambda\mu\nu}\,
Q^\lambda{}_\kappa{}^{\mu\nu}\,.
\end{equation*} Detailed description of the irreducible pieces of curvature and quadratic forms on curvature can be found in \cite{pasic-phd,vassiliev-pseudoinstanton,vassiliev-quadratic}. Our motivation comes from the Yang--Mills theory. The Yang--Mills action
for the affine connection is a special case of (\ref{action}) with
\begin{eqnarray}\label{yang-mills-case}
q(R):=\tensor{R}{^\kappa_{\lambda\mu\nu}}\tensor{R}{^\lambda_{\kappa}^{\mu\nu}}.
\end{eqnarray}

The motivation for choosing a model of gravity which is purely quadratic
in curvature is explained in detail in Section 1 of \cite{vassiliev-quadratic}, Chapter 1 of \cite{pasic-phd} and Section 1 of \cite{pasic-barakovic}.
The idea of using a purely quadratic action in General Relativity
goes back to Hermann Weyl \cite{weyl}, where he argued that the most
natural gravitational action should be quadratic in curvature and
involve all possible invariant quadratic combinations of curvature.
In short, by choosing a purely quadratic curvature
Lagrangian we are hoping to describe phenomena whose characteristic
wavelength is sufficiently small and curvature sufficiently large.
One can get more information and form an idea on the historical development
of the quadratic metric--affine theory of gravity in e.g.
\cite{buchdahl,fairchild1976,fairchild1976erratum,higgs,mielkepseudoparticle,olesen,pasic-phd,pasic-barakovic,pavelleApr1975,stephenson,thompsonFeb1975,thompsonAug1975,vassiliev-quadratic,wilczek,yang}.

\section{Notation}
Our notation follows \cite{King and Vassiliev,pasic-balkanica,pasic-barakovic,pasic-vassiliev,vassiliev-pseudoinstanton,vassiliev-quadratic}. We denote local coordinates by $x^{\mu}$ where
$\mu = 0, 1, 2, 3$, and write $\partial_{\mu}:=\partial/\partial x^{\mu}.$ We define the covariant derivative of a
vector field as
$\displaystyle
\nabla_\mu v^\lambda=\partial_\mu v^\lambda+\tensor{\Gamma}{^{\lambda}_{\mu\nu}}v^{\nu}.
$
The Christoffel symbol
is $\displaystyle \left\{{\lambda\atop {\mu\nu}}\right\}=
\frac{1}{2}g^{\lambda\kappa}(\partial_\mu g_{\nu\kappa}+\partial_\nu g_{\mu\kappa}-\partial_\kappa g_{\mu\nu}).$
We define torsion as
\begin{equation}\label{torsion}
\tensor{T}{^\lambda_{\mu\nu}}=\tensor{\Gamma}{^\lambda_{\mu\nu}}-\tensor{\Gamma}{^\lambda_{\nu\mu}}
\end{equation}
and contortion as
\begin{equation}\label{contortion}
\tensor{K}{^{\lambda}_{\mu\nu}}=\frac{1}{2}\left(
\tensor{T}{^\lambda_{\mu\nu}}+\tensor{T}{_\mu^\lambda_\nu}+\tensor{T}{_\nu^\lambda_\mu}
\right).
\end{equation}
Torsion and contortion are also related as
\begin{equation}\label{torsion-contorsion}
\tensor{T}{^\lambda_{\mu\nu}}=\tensor{K}{^\lambda_{\mu\nu}}-\tensor{K}{^\lambda_{\nu\mu}}.
\end{equation}
The irreducible pieces of torsion are, following \cite{vassiliev-pseudoinstanton},
\begin{equation}\label{dijelovi-torzije1}
T^{(1)}=T-T^{(2)}-T^{(3)},\quad \tensor{{T^{(2)}}}{_{\lambda\mu\nu}}= g_{\lambda\mu}v_\nu-g_{\lambda\nu}v_\mu, \quad
T^{(3)}=\ast w,
\end{equation}
where
\begin{equation}\label{za-torziju}
v_\nu=\frac{1}{3}\tensor{T}{^{\lambda}_{\lambda\nu}},\ \ w_\nu=\frac{1}{6}\sqrt{|\det g|}\tensor{T}{^{\kappa\lambda\mu}}\tensor{\varepsilon}{_{\kappa\lambda\mu\nu}}.
\end{equation}
The pieces $T^{(1)},T^{(2)}$ i $T^{(3)}$ are called \emph{tensor torsion}, \emph{trace torsion}, and \emph{axial torsion}
respectively.
Substituting formulae (\ref{dijelovi-torzije1}) into formula (\ref{contortion}), and formula (\ref{torsion-contorsion}) into formulae
(\ref{za-torziju}) we obtain the irreducible decomposition of contortion:
\begin{equation}\label{contortion1}
K^{(1)}=K-K^{(2)}-K^{(3)}, \quad
\tensor{{K^{(2)}}}{_{\lambda\mu\nu}}=g_{\lambda\mu}v_\nu-g_{\nu\mu}v_\lambda,\quad
K^{(3)}=\frac{1}{2}\ast w,
\end{equation}

where
$\displaystyle
v_\nu=\frac{1}{3}\tensor{K}{^{\lambda}_{\lambda\nu}},\ \ w_\nu=\frac{1}{3}\sqrt{|\det g|}\tensor{K}{^{\kappa\lambda\mu}}\tensor{\varepsilon}{_{\kappa\lambda\mu\nu}}.
$
The irreducible pieces of torsion (\ref{dijelovi-torzije1}) and contortion (\ref{contortion1}) are related as
$ \displaystyle
\tensor{{T^{(i)}}}{_{\kappa\lambda\mu}}=\tensor{{K^{(i)}}}{_{\lambda\kappa\mu}}\ (i=1,2), \
\tensor{{T^{(3)}}}{_{\kappa\lambda\mu}}=2\tensor{{K^{(3)}}}{_{\kappa\lambda\mu}}.$
We define curvature as
$\displaystyle
\tensor{R}{^{\kappa}_{\lambda\mu\nu}}:=\partial_\mu\tensor{\Gamma}{^\kappa_{\nu\lambda}}
-\partial_\nu\tensor{\Gamma}{^\kappa_{\mu\lambda}}
+\tensor{\Gamma}{^\kappa_{\mu\eta}}\tensor{\Gamma}{^\eta_{\nu\lambda}}
-\tensor{\Gamma}{^\kappa_{\nu\eta}}\tensor{\Gamma}{^\eta_{\mu\lambda}},
$
Ricci curvature as $Ric_{\lambda\nu}=\tensor{R}{^{\kappa}_{\lambda\kappa\nu}}$, scalar curvature as $\mathcal{R}=\tensor{Ric}{^\kappa_\kappa}$ and trace-free Ricci
curvature as $\mathcal{R}ic=Ric-\frac{1}{4}\mathcal{R}g.$ We denote Weyl curvature by $\mathcal{W}$ which is understood as the irreducible piece of curvature defined by conditions
$
\tensor{R}{_{\kappa\lambda\mu\nu}}=\tensor{R}{_{\mu\nu\kappa\lambda}}$,
$\tensor{\varepsilon}{^{\kappa\lambda\mu\nu}}\tensor{R}{_{\kappa\lambda\mu\nu}}=0$ and $
Ric=0.
$
We employ the standard convention of raising and lowering tensor indices by means
of the metric tensor. We define the action of the Hodge star on a rank $q$ antisymmetric
tensor as
$\displaystyle
(*Q)_{\mu_{q+1}\ldots\mu_4}\!:=(q!)^{-1}\,\sqrt{|\det g|} \
Q^{\mu_1\ldots\mu_q}\varepsilon_{\mu_1\ldots\mu_4}\,,
$
where $\varepsilon$ is the totally antisymmetric quantity,
$\varepsilon_{0123}:=+1$. When we apply the Hodge star to curvature we have a choice between
acting either on the first or the second pair of indices, so we
introduce two different Hodge stars: the left Hodge star
$\displaystyle ({}^*\!R)_{\kappa\lambda\mu\nu}:=
\frac12\,\sqrt{|\det g|} \ R^{\kappa'\lambda'}{}_{\mu\nu}\,
\varepsilon_{\kappa'\lambda'\kappa\lambda}
$ and the right Hodge star $\displaystyle(R^*)_{\kappa\lambda\mu\nu}:=
\frac12\,\sqrt{|\det g|} \ R_{\kappa\lambda}{}^{\mu'\nu'}\,
\varepsilon_{\mu'\nu'\mu\nu}\,.
$ Given a scalar function $f:M\rightarrow R$ we write for brevity
$$\int f:=\int f \sqrt{\vert \mathrm{det} g \vert}\mathrm{d}x^0\mathrm{d}x^1\mathrm{d}x^2\mathrm{d}x^3,\ \ \mathrm{det} g:=\mathrm{det}(g_{\mu\nu}).$$
\section{Explicit representation of the field equations}\label{main-result}
We write down explicitly our field equations (\ref{variation-S-metric}), (\ref{variation-S-connecion}) under the following assumptions:
\begin{enumerate}
  \item[(i)] our spacetime is metric compatible;
  \item[(ii)] curvature has symmetries \
  $
  \tensor{R}{_{\kappa\lambda\mu\nu}}=\tensor{R}{_{\mu\nu\kappa\lambda}}$ and $\tensor{\varepsilon}{
  ^{\kappa\lambda\mu\nu}}\tensor{R}{_{\kappa\lambda\mu\nu}}=0;
  $
  \item[(iii)] scalar curvature is zero.
\end{enumerate}
The main result of this paper is the following
\begin{theorem}\label{main-result-lemma}
Under the assumptions (i)-(iii) the field equations (\ref{variation-S-metric}) and (\ref{variation-S-connecion}) become
\begin{align}\label{new-equation-variaton-metric}
d_1\tensor{\mathcal{W}}{^{\kappa\lambda\mu\nu}}Ric_{\kappa\mu}+
d_3\left(Ric^{\lambda\kappa}\tensor{Ric}{_{\kappa}^{\nu}}
-\frac{1}{4}g^{\lambda\nu}Ric_{\kappa\mu}Ric^{\kappa\mu}
\right)=0
\end{align}
\begin{align}\label{new-equation-variaton-connection}
 & d_6\nabla_\lambda Ric_{\kappa\mu}-d_7\nabla_\kappa Ric_{\lambda\mu}\\
\nonumber
&+d_6\left(
\tensor{Ric}{_\kappa^\eta}(K_{\mu\eta\lambda}-K_{\mu\lambda\eta})
+\frac{1}{2}g_{\lambda\mu}\tensor{{\mathcal{W}}}{^{\eta\zeta}_{\kappa\xi}}(\tensor{K}{^\xi_{\eta\zeta}}
-\tensor{K}{^\xi_{\zeta\eta}})
+\frac{1}{2}g_{\mu\lambda}\tensor{Ric}{_\xi^\eta}\tensor{K}{^\xi_{\eta\kappa}}
\right.\\
\nonumber
&\quad\quad\quad \left. + g_{\mu\lambda}\tensor{Ric}{_\kappa^\eta}\tensor{K}{^\xi_{\xi\eta}}
-\tensor{K}{^\xi_{\xi\lambda}}Ric_{\kappa\mu}
+\frac{1}{2}g_{\mu\lambda}\tensor{Ric}{_\kappa^\xi}(\tensor{K}{^\eta_{\xi\eta}}-\tensor{K}{^\eta_{\eta\xi}})
\right)\\
\nonumber
&-d_7\left(
\tensor{Ric}{_\lambda^\eta}(K_{\mu\eta\kappa}-K_{\mu\kappa\eta})
+\frac{1}{2}g_{\kappa\mu}\tensor{{\mathcal{W}}}{^{\eta\zeta}_{\lambda\xi}}(\tensor{K}{^\xi_{\eta\zeta}}
-\tensor{K}{^\xi_{\zeta\eta}})
+\frac{1}{2}g_{\mu\kappa}\tensor{Ric}{_\xi^\eta}\tensor{K}{^\xi_{\eta\lambda}}
\right.\\
\nonumber
&\quad\quad\quad\left.-g_{\kappa\mu}\tensor{Ric}{_{\lambda}^{\eta}}\tensor{K}{^\xi_{\xi\eta}}
-\tensor{K}{^\xi_{\xi\kappa}}Ric_{\lambda\mu}+
\frac{1}{2}g_{\mu\kappa}\tensor{Ric}{_\lambda^\xi}(\tensor{K}{^\eta_{\xi\eta}}-\tensor{K}{^\eta_{\eta\xi}})
\right)\\
\nonumber
&+b_{10}\left(
g_{\mu\lambda}\tensor{{\mathcal{W}}}{^{\eta\zeta}_{\kappa\xi}}(\tensor{K}{^\xi_{\zeta\eta}}-\tensor{K}{^\xi_{\eta\zeta}})
+g_{\mu\kappa}\tensor{{\mathcal{W}}}{^{\eta\zeta}_{\lambda\xi}}(\tensor{K}{^\xi_{\eta\zeta}}-\tensor{K}{^\xi_{\zeta\eta}})
\right.\\
\nonumber
&\quad\quad\quad\left.
+g_{\mu\lambda}\tensor{Ric}{_\kappa^\xi}(\tensor{K}{^\eta_{\eta\xi}}-\tensor{K}{^\eta_{\xi\eta}})
+g_{\mu\kappa}\tensor{Ric}{_\lambda^\xi}(\tensor{K}{^\eta_{\xi\eta}}-\tensor{K}{^\eta_{\eta\xi}})\right.\\
\nonumber
&\quad\quad\quad\left.
+g_{\kappa\mu}\tensor{Ric}{_\lambda^\eta}\tensor{K}{^\xi_{\xi\eta}}
-g_{\lambda\mu}\tensor{Ric}{_\kappa^\eta}\tensor{K}{^\xi_{\xi\eta}}
+Ric_{\mu\kappa}\tensor{K}{^\eta_{\lambda\eta}}-Ric_{\mu\lambda}\tensor{K}{^\eta_{\kappa\eta}}\right)\\
\nonumber
&
+2b_{10}\left(\tensor{{\mathcal{W}}}{^{\eta}_{\mu\kappa\xi}}(\tensor{K}{^\xi_{\eta\lambda}}-\tensor{K}{^\xi_{\lambda\eta}})
+\tensor{{\mathcal{W}}}{^{\eta}_{\mu\lambda\xi}}(\tensor{K}{^\xi_{\kappa\eta}}-\tensor{K}{^\xi_{\eta\kappa}})\right.
\\ \nonumber
&\quad\quad\quad\left.
-\tensor{{\mathcal{W}}}{^{\eta\xi}_{\kappa\lambda}}K_{\mu\xi\eta}
-\tensor{{\mathcal{W}}}{^{\eta}_{\mu\lambda\kappa}}\tensor{K}{^\xi_{\xi\eta}}
\right)=0,
\end{align}
where
\begin{align*}
d_1&=b_{912}-b_{922}+b_{10},\quad d_3=b_{922}-b_{911},\\
d_6&=b_{912}-b_{911}+b_{10},\quad d_7=b_{912}-b_{922}+b_{10},
\end{align*}
the b's being coefficients defined in \cite{pasic-phd,pasic-vassiliev}.
\end{theorem}
\begin{remark}
Note that, by definition, we have the curvature symmetry $\tensor{R}{_{\kappa\lambda\mu\nu}}=-\tensor{R}{_{\kappa\lambda\nu\mu}}$, and the symmetry $\tensor{R}{_{\kappa\lambda\mu\nu}}=-\tensor{R}{_{\lambda\kappa\mu\nu}}$ is a consequence of metric compatibility.
\end{remark}
\noindent {\bf {\it Proof.}} The LHS of equations (\ref{new-equation-variaton-metric}) and  (\ref{new-equation-variaton-connection}) are respectively the components of
tensors $A$ and $B$ from the formula
$$\delta S=\int (2A^{\lambda\nu}\delta g_{\lambda\nu}+2\tensor{B}{^{\kappa\mu}_{\lambda}}\delta \tensor{\Gamma}{^{\lambda}_{\mu\kappa}}).$$
Here $\delta g$  and $\delta \Gamma$ are the independent variations of the metric and the connection, and $\delta S$ is the resulting variation of the action.
In deriving explicit formulae for tensors $A$ and $B$ we simplified our calculations by
adopting the following argument. Formula for the quadratic form (\ref{qfmain}) can be, under the assumptions (i)-- (iii), rewritten as
\begin{align*}
q(R) &=\sum_{l,m=1}^2 b_{9lm}(\mathcal{S}^{(l)},\mathcal{S}^{(m)})
+b_{10}(R^{(10)},R^{(10)})_{\mathrm{YM}}+\ldots
\\
&=\sum_{l,m=1}^2 b_{9lm}(\mathcal{R}ic^{(l)},\mathcal{R}ic^{(m)})
+b_{10}(R^{(10)},R^{(10)})_{\mathrm{YM}}+\ldots,
\end{align*}
where by $\cdots$ we denote terms which do not contribute to $\delta S$ when we start our variation
using the assumptions (i)--(iii). In accordance with the convention of \cite{vassiliev-quadratic}, put
\begin{align*}
P_-&:=\frac12(\mathcal{R}ic^{(1)}-\mathcal{R}ic^{(2)}),
\\
P_+&:=\frac12(\mathcal{R}ic^{(1)}+\mathcal{R}ic^{(2)})
=\frac12(Ric^{(1)}+Ric^{(2)}).
\end{align*}
Note that in a metric compatible spacetime $\mathcal{R}ic^{(2)}=-\mathcal{R}ic^{(1)}$, hence $P_+=0$ and  $P_-=\mathcal{R}ic$. Our quadratic form can now
be rewritten as
\begin{align}\label{quadraticform11}
q(R)
&=b_{10}(R^{(10)},R^{(10)})_{\mathrm{YM}}
+(b_{911}-2b_{912}+b_{922})(P_-,P_-)\\ \nonumber
&+2(b_{911}-b_{922})(P_-,P_+)+\ldots .
\end{align}
We also provide another version of this formula which is in accordance with the notation
of \cite{vassiliev-pseudoinstanton}, where most of these terms were studied in detail. The equation (\ref{quadraticform11}) can
be rewritten as
\begin{equation}\label{glavna-kvadratna-forma}
q(\!R)
=c_1(R^{(1)},R^{(1)})_{\mathrm{YM}}
+c_3(R^{(3)},R^{(3)})_{\mathrm{YM}}
+2(b_{911}-b_{922})(P_-,P_+)+\ldots
\end{equation}
where
\begin{equation}\label{coefficients}
c_1=-\frac12(b_{911}-2b_{912}+b_{922}),\qquad c_3=b_{10},
\end{equation}
 and the $R^{(j)}$s  are the irreducible pieces of curvature labeled in accordance with  \cite{vassiliev-pseudoinstanton}.
\subsection*{Variation with respect to the connection}
The variation of $\int (R^{(j)},R^{(j)})_{YM}$ was computed in \cite{vassiliev-pseudoinstanton}:
\begin{equation}\label{variations-rj-rj}
\int (R^{(j)},R^{(j)})_{YM}=4\int \left((\delta_{YM}R^{(j)})^{\mu}(\delta\Gamma)_{\mu}\right)
\end{equation}
where
\begin{equation*}\left(\delta_\mathrm{YM}R\right)^\mu:=\frac1{\sqrt{|{\det}g|}}\,
\left(\partial_\nu+\left[\Gamma_\nu\,,\,\cdot\,\right]\right)
\left(\sqrt{|\det g|}\,R^{\mu\nu}\right)
\end{equation*}
is the Yang-- Mills divergence where we hide the Lie algebra indices of curvature by using matrix notation
\begin{equation}\label{hide-indices}
\tensor{{[\Gamma_\xi,\tensor{R}{_{\mu\nu}}]}}{^{\kappa}_{\lambda}}=
\tensor{\Gamma}{^{\kappa}_{\xi\eta}}\tensor{R}{^{\eta}_{\lambda\mu\nu}}
-\tensor{R}{^{\kappa}_{\eta\mu\nu}}\tensor{\Gamma}{^{\eta}_{\xi\lambda}}.
\end{equation}
In our case, because of assumptions (i)--(iii), the curvature has only two irreducible pieces, namely $R^{(1)}$ i $R^{(3)}=\mathcal{W}$, which can be written as
\begin{eqnarray*}
\tensor{{R^{(1)}}}{_{\kappa\lambda\mu\nu}}&=&\frac{1}{2}\left(g_{\kappa\mu}Ric_{\lambda\nu}
-g_{\lambda\mu}Ric_{\kappa\nu}-g_{\kappa\nu}Ric_{\lambda\mu}+g_{\lambda\nu}Ric_{\kappa\mu}\right),\\
\tensor{{R^{(3)}}}{_{\kappa\lambda\mu\nu}}&=&\tensor{R}{_{\kappa\lambda\mu\nu}}-\tensor{{R^{(1)}}}{_{\kappa\lambda\mu\nu}}.
\end{eqnarray*}
with the other $R^{(j)}$s being zero. Substituting these expressions into (\ref{variations-rj-rj})
we get
\begin{align}\label{r1-r1-po-konekciji}
\delta\int & (R^{(1)},R^{(1)})_{YM}= 2\int (\delta\Gamma^{\lambda\mu\kappa})
\left[\nabla_\lambda Ric_{\kappa\mu}
-\nabla_\kappa\tensor{Ric}{_{\lambda\mu}}
+g_{\kappa\mu}\nabla_\xi\tensor{Ric}{_{\lambda}^{\xi}}\right.\\
\nonumber
&-g_{\mu\lambda}\nabla_\xi\tensor{Ric}{_{\kappa}^{\xi}}
+\tensor{Ric}{_{\kappa}^{\eta}}(\tensor{K}{_{\mu\eta\lambda}}-\tensor{K}{_{\mu\lambda\eta}})
+\tensor{Ric}{_{\lambda}^{\eta}}(\tensor{K}{_{\mu\kappa\eta}}-\tensor{K}{_{\mu\eta\kappa}})
\\
\nonumber
&\left.
+g_{\mu\lambda}\tensor{K}{^{\xi}_{\xi\eta}}\tensor{Ric}{_{\kappa}^{\eta}}
-\tensor{K}{^{\eta}_{\eta\lambda}}\tensor{Ric}{_{\kappa\mu}}
+\tensor{K}{^{\xi}_{\xi\kappa}}\tensor{Ric}{_{\lambda\mu}}
-\tensor{g}{_{\kappa\mu}}\tensor{K}{^{\xi}_{\xi\eta}}\tensor{Ric}{_{\lambda}^{\eta}}\right]
\end{align}
and
\begin{equation}\label{r3-r3-po-konekciji}
\delta\!\!\int(R^{(3)},R^{(3)})_{YM}\!=\!4\!\int(\delta\tensor{\Gamma}{^{\lambda\mu\kappa}})(
\nabla_\nu\tensor{\mathcal{W}}{_{\kappa\lambda\mu}^{\nu}}\!
-\tensor{K}{_{\mu\nu\eta}}\tensor{\mathcal{W}}{_{\kappa\lambda}^{\eta\nu}}\!
-\tensor{{K}}{^{\xi}_{\xi\nu}}\tensor{\mathcal{W}}{_{\kappa\lambda\mu}^{\nu}}).
\end{equation}
Further,
\begin{flalign}\label{varijacijapp-po-konekciji}
\delta\int (P_{-},P_{+})=-\frac{1}{2}\int(\delta\Gamma^{\lambda\mu\kappa}) \left[\nabla_\lambda Ric_{\kappa\mu}+\nabla_\kappa\tensor{Ric}{_{\lambda\mu}}-g_{\mu\lambda}\nabla_\xi \tensor{Ric}{_{\kappa}^{\xi}}
\right.\\
\nonumber
-g_{\kappa\mu}\nabla_\xi\tensor{Ric}{_{\lambda}^{\xi}}+\tensor{Ric}{_{\kappa}^{\eta}}(\tensor{K}{_{\mu\eta\lambda}}-\tensor{K}{_{\mu\lambda\eta}})
+\tensor{Ric}{_{\lambda}^{\eta}}(\tensor{K}{_{\mu\eta\kappa}}-\tensor{K}{_{\mu\kappa\eta}})\\
\nonumber
\left.+\tensor{K}{^{\xi}_{\xi\eta}}(g_{\mu\lambda}\tensor{Ric}{_{\kappa}^{\eta}}+g_{\kappa\mu}\tensor{Ric}{_{\lambda}^{\eta}})
-\tensor{K}{^{\xi}_{\xi\lambda}}\tensor{Ric}{_{\kappa\mu}}-\tensor{K}{^{\xi}_{\xi\kappa}}\tensor{Ric}{_{\lambda\mu}}\right].
\end{flalign}
Combining formulae (\ref{glavna-kvadratna-forma}), (\ref{coefficients}), (\ref{r1-r1-po-konekciji})--(\ref{varijacijapp-po-konekciji}), we arrive at the explicit form of the
field equation (\ref{new-equation-variaton-connection}):
\begin{align}
\label{jednacina-polja-po-konekciji-prva}
&d_6'\nabla_{\lambda}Ric_{\kappa\mu}-d_7'\nabla_{\kappa}Ric_{\lambda\mu}+
\\
\nonumber
&d_6'\left(-g_{\mu\lambda}\tensor{{\nabla_{\eta}Ric}}{_{\kappa}^{\eta}}
+\tensor{Ric}{_{\kappa}^{\eta}}(K_{\mu\eta\lambda}-K_{\mu\lambda\eta})
+g_{\mu\lambda}\tensor{K}{^{\xi}_{\xi\eta}}\tensor{Ric}{_{\kappa}^{\eta}}
-\tensor{K}{^{\xi}_{\xi\lambda}}\tensor{Ric}{_{\kappa\mu}}\right)-\\
\nonumber
&d_7'\left(-g_{\kappa\mu}\tensor{{\nabla_{\eta}Ric}}{_{\lambda}^{\eta}}
+\tensor{Ric}{_{\lambda}^{\eta}}(K_{\mu\eta\kappa}-K_{\mu\kappa\eta})
+g_{\mu\kappa}\tensor{K}{^{\xi}_{\xi\eta}}\tensor{Ric}{_{\lambda}^{\eta}}
-\tensor{K}{^{\xi}_{\xi\kappa}}\tensor{Ric}{_{\lambda\mu}}\right)+\\
\nonumber
&2b_{10}\left(\nabla_\eta\tensor{{\mathcal{W}}}{^{\eta}_{\mu\lambda\kappa}}
-K_{\mu\xi\eta}\tensor{{\mathcal{W}}}{^{\eta\xi}_{\kappa\lambda}}
+\tensor{K}{^{\xi}_{\xi\eta}}\tensor{{\mathcal{W}}}{^{\eta}_{\mu\kappa\lambda}}
\right)=0,
\end{align}
where
$$d_6'=b_{912}-b_{911},\ \ d_7'=b_{912}-b_{922}.$$
Let us use the Bianchi identity for curvature
\[
(\partial_\xi+[\Gamma_\xi,\cdot])\tensor{R}{_{\mu\nu}}+(\partial_\nu+[\Gamma_\nu,\cdot])\tensor{R}{_{\xi\mu}}+
(\partial_\mu+[\Gamma_\mu,\cdot])\tensor{R}{_{\nu\xi}}=0,
\]
where we hide the Lie algebra indices of curvature by using formula (\ref{hide-indices}).
Using our assumptions (i)--(iii) and making one contraction of indices, we get
\begin{align}
\label{bjanki22}
&\nabla_\xi Ric_{\lambda\nu}
-\nabla_\nu Ric_{\lambda\xi}+g_{\lambda\xi}\nabla_\mu Ric^\mu{}_\nu
-g_{\lambda\nu}\nabla_\mu Ric^\mu{}_\xi+\\
\nonumber
&
Ric^\mu{}_\eta
(g_{\lambda\xi}K^\eta{}_{\mu\nu}-g_{\lambda\nu}K^\eta{}_{\mu\xi})+Ric^\mu{}_\xi(K_{\nu\mu\lambda}-K_{\mu\nu\lambda})+\\
\nonumber
&
Ric^\mu{}_\nu(K_{\mu\xi\lambda} -K_{\xi\mu\lambda}) +Ric_{\lambda\nu}(K^\mu{}_{\xi\mu} -K^\mu{}_{\mu\xi})+Ric_{\lambda\xi}(K^\mu{}_{\mu\nu} -K^\mu{}_{\nu\mu})+\\
\nonumber
& 2\left(\nabla_\mu \mathcal{W}^\mu{}_{\lambda\nu\xi}
+\mathcal{W}^\mu{}_{\lambda\xi\eta}(K^\eta{}_{\nu\mu} - K^\eta{}_{\mu\nu})
+ \mathcal{W}^\mu{}_{\lambda\nu\eta}(K^\eta{}_{\mu\xi}-K^\eta{}_{\xi\mu})\right)=0.
\end{align}
Another contraction in (\ref{bjanki22}) yields
\begin{equation} \label{nabla-ricc}
\nabla_\eta Ric^\eta{}_\kappa  =
 -\frac{1}{2}Ric^\eta{}_\xi K^\xi{}_{\eta\kappa}
 -\frac{1}{2} \tensor{Ric}{^{\xi}_{\kappa}}(\tensor{K}{^{\eta}_{\xi\eta}}-\tensor{K}{^{\eta}_{\eta\xi}})
 - \frac{1}{2} \mathcal{W}^{\eta\zeta}{}_{\kappa\xi}(K^\xi{}_{\eta\zeta}-K^\xi{}_{\zeta\eta}).
 \end{equation}
Substitution of (\ref{nabla-ricc}) into (\ref{bjanki22}) gives
\begin{align}
\label{nabla-wejl}
\nabla_\eta & \mathcal{W}^\eta{}_{\mu\lambda\kappa} =
\mathcal{W}^\eta{}_{\mu\kappa\xi}(K^\xi{}_{\eta\lambda}-K^\xi{}_{\lambda\eta})
+\mathcal{W}^\eta{}_{\mu\lambda\xi}(K^\xi{}_{\kappa\eta}-K^\xi{}_{\eta\kappa})\\
\nonumber
&+\frac{1}{4}(K^\xi{}_{\zeta\eta}-K^\xi{}_{\eta\zeta})
(g_{\mu\lambda} \mathcal{W}^{\eta\zeta}{}_{\kappa\xi}
-g_{\mu\kappa} \mathcal{W}^{\eta\zeta}{}_{\lambda\xi})\\
\nonumber
&+\frac{1}{2}\left[\nabla_\lambda Ric_{\mu\kappa}
-\nabla_\kappa Ric_{\mu\lambda}
+Ric^\eta{}_\kappa(K_{\eta\lambda\mu}-K_{\lambda\eta\mu})\right.\\
\nonumber
&\left.+Ric^\eta{}_\lambda(K_{\kappa\eta\mu}-K_{\eta\kappa\mu})
+Ric_{\mu\lambda}(K^\eta{}_{\eta\kappa}-K^\eta{}_{\kappa\eta}) +Ric_{\mu\kappa}(K^\eta{}_{\lambda\eta}-K^\eta{}_{\eta\lambda})\right]\\
\nonumber
&+\frac{1}{4}Ric^\eta{}_\xi(g_{\mu\lambda}K^\xi{}_{\eta\kappa} - g_{\mu\kappa}K^\xi{}_{\eta\lambda})
+\frac{1}{4}(\tensor{K}{^{\eta}_{\eta\xi}}-\tensor{K}{^{\eta}_{\xi\eta}})
(g_{\mu\lambda}\tensor{Ric}{^{\xi}_{\kappa}}-g_{\mu\kappa}\tensor{Ric}{^{\xi}_{\lambda}}).
\end{align}
Formulae (\ref{nabla-ricc}) and (\ref{nabla-wejl}) allow us to exclude the terms with
$\nabla_\eta\tensor{Ric}{_{\kappa}^{\eta}}$, $\nabla_\eta\tensor{Ric}{_{\lambda}^{\eta}}$ i $\nabla_\eta\tensor{{\mathcal{W}}}{^{\eta}_{\mu\lambda\kappa}}$ from equation (\ref{jednacina-polja-po-konekciji-prva}), reducing the latter to (\ref{new-equation-variaton-connection}).

\subsection*{Variation with respect to the metric} The field equation (\ref{new-equation-variaton-metric}) is identical to the
one in the Riemannian case as given in \cite{vassiliev-quadratic}, only with the scalar curvature being zero, which is not surprising as the assumptions on the torsion do not influence the form of the equation. Here we present briefly the derivation of equation (\ref{new-equation-variaton-metric}).  A lengthy but straightforward calculation shows that
\begin{align}\label{r1-r1-po-metrici}
\delta\int \left(R^{(1)},R^{(1)}\right)_{YM}&=-2\int\tensor{\mathcal{W}}{^{\kappa\beta\alpha\nu}}Ric_{\kappa\nu}\delta g_{\alpha\beta}.\\
\label{r3-r3-po-metrici}
\delta\int \left(R^{(3)},R^{(3)}\right)_{YM}&=-2\int\tensor{\mathcal{W}}{^{\kappa\beta\alpha\nu}}Ric_{\kappa\nu}\delta g_{\alpha\beta}.
\end{align}
Further,
\begin{eqnarray}\label{varijacijapp-po-metrici}
\delta\int(P_{-},P_{+})=\int(Ric,\delta P_{+})=\frac{1}{2}\int(Ric,\delta Ric)+\frac{1}{2}\int(Ric,\delta Ric^{(2)})=\\
\nonumber
-\frac{1}{4}\int \left(4Ric^{\kappa\alpha}\tensor{Ric}{_{\kappa}^{\beta}}
+2\tensor{\mathcal{W}}{^{\kappa\alpha\beta\nu}}\tensor{Ric}{_{\kappa\nu}} -g^{\alpha\beta}\tensor{Ric}{_{\mu\nu}}\tensor{Ric}{^{\mu\nu}}\right)\delta g_{\alpha\beta}.
\end{eqnarray}
Combining formulae (\ref{glavna-kvadratna-forma}) - (\ref{varijacijapp-po-metrici}) we arrive at the explicit form of the
field equation (\ref{new-equation-variaton-metric}). \newline This ends the proof of Theorem~\ref{main-result-lemma}. \qed

\begin{remark}\label{purelytensor}
If we assume that torsion is purely tensor, in addition to our assumptions (i)-(iii),
the field equations (\ref{new-equation-variaton-metric}), (\ref{new-equation-variaton-connection}) reduce to those presented in \cite{pasic-phd,pasic-vassiliev},
with the correction presented in Appendix C of \cite{pasic-barakovic}.
\end{remark}
\begin{remark}\label{purelyaxial}
If we assume that the torsion is purely axial, in addition to our assumptions (i)-(iii),
the field equations (\ref{new-equation-variaton-metric}), (\ref{new-equation-variaton-connection}) reduce to
\begin{align}\label{prvaAxial}
d_1\tensor{\mathcal{W}}{^{\kappa\lambda\mu\nu}}Ric_{\kappa\mu}+
d_3\left(Ric^{\lambda\kappa}\tensor{Ric}{_{\kappa}^{\nu}}
-\frac{1}{4}g^{\lambda\nu}Ric_{\kappa\mu}Ric^{\kappa\mu}
\right)&=0, \\
\label{drugaAxial}
d_6\left(\nabla_\lambda  Ric_{\kappa\mu} + \tensor{Ric}{_\kappa^\eta}T_{\mu\eta\lambda}
\right)
- d_7\left( \nabla_\kappa Ric_{\lambda\mu} +
\tensor{Ric}{_\lambda^\eta}T_{\mu\eta\kappa}
\right) &\\
\nonumber
+2b_{10}
\left(\tensor{{\mathcal{W}}}{^{\eta}_{\mu\kappa\xi}}\tensor{T}{^\xi_{\eta\lambda}}
+\tensor{{\mathcal{W}}}{^{\eta}_{\mu\lambda\xi}}\tensor{T}{^\xi_{\kappa\eta}} -\frac12\tensor{{\mathcal{W}}}{^{\eta\xi}_{\kappa\lambda}}T_{\mu\xi\eta}
\right)&=0.
\end{align}
where the d's are the same as given in Theorem \ref{main-result-lemma} and the b's are the same as given in \cite{pasic-phd, pasic-vassiliev}.
\end{remark}
\begin{remark}
An effective technique for writing down the field equations explicitly can be found in \cite{tele3, hehlreview}. Namely, according to formulae (142), (143) of \cite{tele3}, our system of field equations reads
\begin{eqnarray}
\label{anholonomic1}
e_\alpha \rfloor V - (e_\alpha \rfloor R_{\beta}{}^\gamma \wedge \frac{\partial V}{\partial R_\beta{}^\gamma}) & = & 0, \\
\label{anholonomic2}
D \frac{\partial V}{\partial R_\alpha{}^\beta} & = & 0.
\end{eqnarray} Here the notation is \emph{anholonomic},
$V := *q(R)$ is the Lagrangian, $e_\alpha$ is the frame and $D$ is the covariant exterior differential, $\rfloor$ is the interior product and the exterior product is $\wedge$. Equation (\ref{anholonomic2}) is the explicit form of equation (\ref{variation-S-connecion}), but equations (\ref{anholonomic1}) and (\ref{variation-S-metric}) are somewhat different: the difference is that (\ref{anholonomic1}) is the result of variation with respect to the frame rather than the metric. It is known, however, that the systems (\ref{variation-S-metric}), (\ref{variation-S-connecion}) and (\ref{anholonomic1}), (\ref{anholonomic2}) are equivalent.
\end{remark}

\section{Discussion}

A comprehensive study of equations (\ref{variation-S-metric}), (\ref{variation-S-connecion}) was done only relatively recently. Vassiliev \cite{vassiliev-quadratic} solved the problem of existence and
uniqueness for Riemannian solutions (see Definition \ref{riemann}). He showed that the
Riemannian solutions of the equations
(\ref{variation-S-metric}), (\ref{variation-S-connecion}) are Einstein spaces, pp-waves with parallel Ricci curvature and Riemannian spacetimes which have zero scalar curvature and are locally a product of Einstein 2--manifolds. \newline
Furthermore, in the same paper \cite{vassiliev-quadratic} Vassiliev showed that
the above spacetimes are the \emph{only} Riemannian solutions of the
system of field equations (\ref{variation-S-metric}), (\ref{variation-S-connecion}). It is
also interesting that before \cite{vassiliev-quadratic} it had not been noticed that
pp-waves were solutions of the problem, although they were well
known spacetimes in theoretical physics. Because of the uniqueness result
we can now only establish new non-Riemannian solutions of
the system (\ref{variation-S-metric}), (\ref{variation-S-connecion}).

In \cite{vassiliev-quadratic} Vassiliev also presented one non-Riemannian
solution of the system (\ref{variation-S-metric}), (\ref{variation-S-connecion}) and it
was a \emph{torsion wave} solution with explicitly given torsion. For the Yang--Mills case (\ref{yang-mills-case}) this torsion wave solution was first obtained by Singh and Griffiths: see last
paragraph of Section 5 in \cite{griffiths3} and the same solution was later independently rediscovered by
King and Vassiliev in \cite{King and Vassiliev}.
It should be pointed out that
the torsion wave solution of King and Vassiliev
is a highly specialised version of the solution obtained by Singh
and Griffiths \cite{griffiths3}, which is a solution of algebraic type III,
where the Riemannian spacetime is a Kundt plane-fronted gravitational wave
and the torsion is purely tensor.
Vassiliev's contribution in \cite{vassiliev-quadratic} was to show that these spacetimes
satisfy equations (\ref{variation-S-metric}), (\ref{variation-S-connecion}) in the most general case of the
purely quadratic action (\ref{action}). This work of Vassiliev went on to conclude that this torsion wave
was a non-Riemannian analogue of a \emph{pp-wave}, whence came the
motivation for generalising the notion of a classical Riemannian pp-wave to spacetimes
with torsion in such a way as to incorporate the non-Riemannian torsion-wave
solution into the construction.

PP-waves are well known spacetimes in general relativity, first
discovered by Brinkmann \cite{brinkmann} in 1923, and subsequently
rediscovered by several authors, for example Peres \cite{peres} in 1959.
We define a \emph{pp-wave} as a Riemannian spacetime which admits a
nonvanishing parallel spinor field, or equivalently as a Riemannian spacetime
whose metric can be written locally in the form
$\displaystyle
\rmd s^2= \,2\,\rmd x^0\,\rmd x^3-(\rmd x^1)^2-(\rmd x^2)^2
+f(x^1,x^2,x^3)\,(\rmd x^3)^2
$
in some local coordinates $(x^0,x^1,x^2,x^3)$.  In our previous work \cite{pasic-phd,pasic-balkanica,pasic-barakovic,pasic-vassiliev}, where a detailed description of pp-waves can be found, we presented results which were new (non-Riemannian) explicit vacuum solutions of the system of our field equations (\ref{variation-S-metric}), (\ref{variation-S-connecion}), namely generalised pp-waves with torsion. This generalisation was done by employing the pp-metric and giving an explicit torsion, identical to the torsion-wave obtained by Vassiliev in \cite{vassiliev-quadratic}. The fact that the two solutions, one Riemannian and the other non-Riemannian, `add up' is extremely non-trivial, as the system we are observing is highly non-linear. We further explored the properties and characteristics of these generalised pp-waves, showing that they are indeed solutions of the system of our field equations (\ref{variation-S-metric}), (\ref{variation-S-connecion}), by writing the field equations explicitly, like in this paper, but with the additional assumption of torsion being purely tensor, which simplifies matters substantially. Our analysis of vacuum solutions of QMAG showed that classical pp-spaces of parallel Ricci curvature should not be viewed on their own, but that they are in fact
a particular (degenerate) representative of a wider class of solutions, namely, generalised pp-spaces of parallel Ricci curvature. The latter appear to admit a sensible physical interpretation, which we explored in detail in \cite{pasic-barakovic} where we gave a comparison with the classical model describing the interaction of gravitational and massless neutrino fields, namely Einstein--Weyl theory, constructed pp-wave type solutions of this theory and pointed out that generalised pp-waves of parallel Ricci curvature are very similar to pp-wave type solutions of the Einstein--Weyl model. Therefore we proposed that our generalised pp-waves of parallel Ricci curvature represent a metric-affine (i.e. conformally invariant) model for a massless particle, namely the \emph{massless neutrino}. The main difference in using our metric--afine model is
that Einstein--Maxwell and Einstein--Weyl theories contain the gravitational constant which dictates a particular relationship between the strengths of the fields in question, whereas our model
is conformally invariant and the amplitudes of the two curvatures (i.e. torsion generated and metric generated curvatures) are totally independent.

The main idea of the current paper is to empower us to find new pp-wave type non-Riemannian solutions with \emph{arbitrary} (as opposed to purely tensor) torsion. Note that the assumptions (i)-(iii) used to derive
our equations are automatically satisfied by pp-waves and their generalisation, so we are justified in using them.

The observation that one can construct vacuum solutions of QMAG in terms of pp-waves is a recent development.
The fact that classical pp-waves of parallel Ricci curvature are
solutions was first pointed out in \cite{garda,poland,vassiliev-quadratic}.
There are a number of publications in which authors suggested
various generalisations of the concept of a classical pp-wave. These
generalisations were performed within the Riemannian setting and usually involved the
incorporation of a constant non-zero scalar curvature; see
\cite{obukhov pp} and extensive further references therein. Our
construction in \cite{pasic-phd,pasic-balkanica,pasic-barakovic,pasic-vassiliev}
generalised the concept of a classical pp-wave in a different
direction: we added torsion while retaining zero scalar curvature. Note that we keep this assumption in the current paper.

A powerful method which in the past has been used for the
construction of vacuum solutions of QMAG
is the so-called \emph{double duality ansatz}
\cite{baekleretal1,baekleretal2,mielkepseudoparticle,mielkeduality,vassiliev-pseudoinstanton,vassiliev-quadratic}.
For certain types of
quadratic actions the following is known to be
true: if the spacetime is metric compatible and curvature is
irreducible (i.e. all irreducible pieces except one are identically
zero) then this spacetime is a solution of
(\ref{variation-S-metric}), (\ref{variation-S-connecion}). This
fact is referred to as the double duality ansatz because the proof
is based on the use of the double duality transform
$R\mapsto{}^*\!R^*$ (this idea is due to Mielke
\cite{mielkepseudoparticle}) and because the above conditions imply
${}^*\!R^*=\pm R$. However, solutions presented in \cite{pasic-phd,pasic-balkanica,pasic-barakovic,pasic-vassiliev} and the ones we hope arise from our current work do not fit into the double duality scheme. This is due to reasons that can be found explained in detail in \cite{pasic-phd,pasic-vassiliev}. These solutions are similar to those of
Singh and Griffiths \cite{griffiths3}. The main differences are as
follows:
\begin{itemize}
\item
The solutions in \cite{griffiths3} satisfy the condition
$\{\!Ric\!\}=0$ whereas our solutions satisfy the weaker condition
$\{\!\nabla\!\}\{\!Ric\!\}=0$.
\item
The solutions in \cite{griffiths3} were obtained for the Yang--Mills
case (\ref{yang-mills-case}) whereas we deal with a general
$\mathrm{O}(1,3)$-invariant quadratic form $q$ with 16 coupling
constants.
\end{itemize}
One interesting generalisation of the concept of a pp-wave was
presented by Obukhov in \cite{obukhov new}. Obukhov's
motivation comes from his previous work \cite{obukhov pp} which is
the Riemannian case. In fact, the ansatz for the metric and the
coframe of \cite{obukhov new} is exactly the same as in the
Riemannian case. However, the connection extends the Levi-Civita
connection is such a way that torsion and \emph{nonmetricity}
($\nabla g \ne 0$) are present, and are determined by this
extension of the connection. Obukhov studies the same general quadratic Lagrangian with 16 terms,
and the result of \cite{obukhov new} does not belong to the triplet ansatz, see
\cite{hehlandmaciasexactsolutions2, obukhov triplet}. Obukhov's gravitational wave solutions have only two
non-trivial pieces of curvature. However, unlike in our setting,
the two non-zero pieces of curvature in \cite{obukhov new} are equivalent
to the pieces of curvature coming from the $10$-dimensional $\mathbf{R}^{(10)}$
and the $30$-dimensional $\mathbf{R}^{(30)}$ irreducible curvature
subspaces. Hence the main differences between our work and Obukhov's
generalisation of \cite{obukhov new} are the following:
\begin{itemize}
\item In Obukhov's plane-fronted waves not only are the torsion waves present,
but the non-metricity has a non-trivial wave behaviour as well. As we are
only looking at metric--compatible spacetimes,
nonmetricity cannot appear in our construction.
\item The second ($\mathbf{R}^{(30)}$) irreducible piece of curvature cannot appear in our ansatz, as this piece of curvature is zero for metric-compatible spacetimes.
\item Obukhov's gravitational wave
solutions provide a minimal generalisation of the
pseudoinstanton, see \cite{vassiliev-pseudoinstanton}, in the sense that nonmetricity does not vanish
and that curvature has two non-zero pieces.
\end{itemize}
In relation to our goal of finding new solutions of QMAG, the
two papers of Singh \cite{singh-axial, singh-trace} are of special interest to us. Singh
constructs solutions for the Yang--Mills case (\ref{yang-mills-case}) with purely axial and purely trace torsion respectively and unlike the solution of \cite{griffiths3},
$\{\!Ric\!\}$ is not assumed to be zero. It is obvious that these solutions differ
from the ones presented in \cite{pasic-vassiliev}, as the torsion there is assumed to be purely tensor and the torsion wave produced curvature is purely Weyl, i.e. $\{\!Ric\!\}=0$.
It would however be of interest to us to see whether this construction of Singh's can be expanded
to our most general $\mathrm{O}(1,3)$-invariant quadratic form $q$ with 16 coupling
constants. In \cite{singh-axial} Singh presents solutions of the field equations (\ref{variation-S-metric}), (\ref{variation-S-connecion}) for the Yang--Mills case (\ref{yang-mills-case}) for a purely
\emph{axial} torsion. The make a class of solutions that cannot be obtained using the double duality ansatz,
see \cite{baekleretal1,baekleretal2,mielkepseudoparticle,mielkeduality,vassiliev-pseudoinstanton,vassiliev-quadratic}.
In fact, Singh uses the `spin coefficient technique' from his previous work with Griffiths \cite{griffiths3} in constructing the new solutions.
In view of the fact that the previous purely tensor solutions in \cite{griffiths3} were shown to also be the solutions in the most general case (\ref{qfmain}), we expect that this is also true in the purely
axial case. Therefore, similarly to our previously found purely tensor torsion waves, we suggest the following
\begin{conjecture}
There exist purely axial torsion waves which are solutions of the field equations (\ref{variation-S-metric}), (\ref{variation-S-connecion}).
\end{conjecture} We should point out that the explicit forms (\ref{new-equation-variaton-metric}), (\ref{new-equation-variaton-connection}) of our fields equations (\ref{variation-S-metric}), (\ref{variation-S-connecion}) given in Section \ref{main-result} were obtained without any a priori assumptions
on torsion. Hence, under the assumption of purely axial torsion, the field equations would be
substantially simplified, as given in equations (\ref{prvaAxial}), (\ref{drugaAxial}) in Remark \ref{purelyaxial}.

Following the reasoning behind the generalised pp-waves of \cite{pasic-vassiliev} that were shown to be solutions of the field equations (\ref{variation-S-metric}), (\ref{variation-S-connecion}), where we `combined' the pp-metric and the purely tensor torsion waves to obtain a new class of solutions for QMAG, we hope to be able to do the same with purely axial torsion waves. Therefore, we suggest the following
\begin{conjecture}
There exists a class of spacetimes equipped with the pp-metric and explicitly given purely axial torsion of parallel Ricci curvature that satisfies the field equations (\ref{variation-S-metric}), (\ref{variation-S-connecion}).
\end{conjecture}
Similarly, in \cite{singh-trace} Singh presents solutions of the field equations (\ref{variation-S-metric}), (\ref{variation-S-connecion}) for the Yang--Mills case (\ref{yang-mills-case}) for a purely
\emph{trace} torsion. At this point we are still not sure whether these torsion waves can also be used to create new generalised pp-wave solutions of QMAG, but we hope to be able to answer this question as well, together with proving the two conjectures above, which would accomplish the main purpose of the current article -- to make it simpler for researchers to find and confirm new solutions of metric--affine gravity. The next step would then be to give a physical interpretation of these new solutions by comparing them to existing Riemannian solutions, like it was done in \cite{pasic-barakovic} for purely tensor torsion generalised pp-waves, which would represent a very valuable scientific contribution in the field of alternative theories of gravity.
\section*{Acknowlegements}
The authors are very grateful to Dmitri Vassiliev for valuable advice. This work was partially supported by the grant of the Ministry of Education and Science of the Federation of Bosnia and Herzegovina.

\label{lastpage}

\end{document}